# Towards a simple, comprehensive model of regular earthquakes and slow slip events, part I: one-dimensional model

Naum I. Gershenzon*[1], Cong Zhou[2], and Thomas Skinner[1]

[1]Physics Department, Wright State University, 3640 Colonel Glenn Highway Dayton, OH 45435

[2]The Second Monitoring and Application Center, No.316 Xiying Road, Xi'an City, Shaanxi Province, China, 710054

*Corresponding author

E-mail address: naum.gershenzon@wright.edu

**Abstract**

Seismic events are a prominent manifestation of the earth's fault dynamics. Since the focal area of these events is inaccessible to direct observation, modeling plays a vital role in the investigation of processes below the earth's surface. We have developed a model that describes the major characteristics of a rupture, ranging from regular earthquakes (EQs) to slow slip events (SSEs), including episodic tremor and slip (ETS). Previous model predictions, while accurate, are based on a highly idealized initial stress distribution and a simple velocity-dependent expression for friction. The full scope of the model has, therefore, not been fully demonstrated. Further developments, presented here, include more physically realistic treatments of both the initial conditions and friction. Important new model predictions are: (1) The type of a seismic event, i.e. regular EQ or SSE, is determined by the fault strength, the shear to normal stress ratio, and the gradient in the ratio. Quantitative values for these crucial parameters are also obtained here. The critical factor for the appearance of a regular EQ is the abrupt decrease of friction at slip velocities $\geq 0.1$ m/s due to flash heating or other mechanisms. If the stress ratio and spatial gradient of this ratio are large enough to accelerate slip velocity up to ~0.1 m/s, then a regular EQ occurs. For lower values, all types of SSEs are possible; (2) Rupture velocities for regular EQs range from a fraction of the shear wave velocity, $c_s$, up to the supershear velocity. The range for SSEs is much wider, from $10^{-7}c_s$ to $10^{-1}c_s$. The maximum slip velocity for regular EQs is typically on the order of 1 m/s. For SSEs, the slip velocity ranges widely from a few cm/year up to 0.1 m/s. The magnitude of the stress drop may vary in the range from 1% to 10% of the initial shear stress for regular EQs and from 0.1% to 10% for the SSEs. Generally, the stress drop in SSEs is one to three orders of magnitude less than in EQs; (3) The rupture can expend as a crack-like mode or as a self-healing pulse mode. The type of rupture mode is determined by the stress ratio and its gradient. If stress heterogeneity has a step-like shape, the rupture is always crack-like. If the heterogeneity is localized, the rupture is crack-like for sufficiently large values of the stress and its gradient (as quantified in this work) and is pulse-like for smaller values. The type of heterogeneity also determines the position of maximum slip; (4) After an instability develops, rupture dynamics do not depend on the relative values of the rate-state *a* and *b* parameters, i.e., it does not matter if frictional sliding is velocity-weakening (*a* < *b*) or velocity-strengthening (*a* > *b*). Popular models based on the rate-state frictional law are not consistent with this result.

## 1. Introduction

The ultimate goal of earthquake research is a predictive theory for seismic events. Towards that end, we have developed a model and investigated its relevance for studying the dynamics of earth's crustal faults [1-8]. The model is derived solely from physical principles, has no adjustable parameters, and is comparatively simple to use. In this article, we present further developments in the model and evaluate its new predictions compared to observed phenomena.

Seismic events can be arbitrarily divided into two main categories: regular earthquakes (EQs) and slow slip events (SSEs). They result from shear stress relaxation accumulated on the boundaries of the earth's faults. Other common terms for SSEs are 'slow EQs', 'silent EQs', 'aseismic strain transients', and 'creep events.' Regular EQs have obviously been known for a long time [9], but SSEs were discovered comparatively recently [10-12] and are now widely observed along major transform and subduction zones [13-19]. Although large regular EQs are the most intriguing part of fault dynamics, SSEs are viewed as increasingly important, since a large fraction of plate motion and energy release is due to aseismic movement along a fault [20]. The various mechanisms of stress relaxation therefore need to be understood in greater detail [21].

The similarities and differences between regular EQs and SSEs provide important observational constraints on any theory describing their underlying mechanisms. Both release substantial amounts of accumulated strain energy by spontaneously initiated slip along a fault. However, the latter does not radiate seismic waves. Whereas the rupture propagation velocity of EQs is about 3 km/s and the slip velocity is about 1 m/s, the propagation speed of SSEs is less by up to five orders of magnitude and the slip velocity is less by up to ten orders of magnitude. The scaling laws are also different [22-24].

The focal area of an EQ/SSE is usually located at depths inaccessible to direct observation. Seismic waves and/or transient surface deformation, observed on the Earth's surface or at shallow depth, are the only manifestations of seismic events. Consequently, modeling is especially important for understanding processes at their source. However, the phenomenology underlying seismic events is sufficiently complicated that there are no governing equations that adequately describe crust rheology, composition, and structure, as well as the highly nonlinear character of frictional sliding along faults [25]. Therefore, various approaches have been developed to construct models for specific purposes.

The early models had exact analytical solutions that were applied to the dynamics of a crack in elastic solids. This approach revealed the fundamental relations between kinematic variables (rupture velocity, slip velocity, displacement) and dynamic variables (shear and normal stress, stress drop) characterizing EQs [26-30].

The limitations of purely analytical models were subsequently elucidated [28], leading to the development of computational models that solve a system of elasto-dynamic equations with boundary conditions given as a frictional law on crack boundaries, e.g. [31-37]. Whereas these essentially 3D models are able to describe some basic characteristics of an earthquake and accompanying seismic radiation, they are beset by various challenges, both mathematical (singularities at the tip of the crack; accuracy and relevance of the chosen boundary conditions, i.e., fault constitutive relations) and computational (issues of convergence, calculational load, and efficiency). Concurrently, a simpler 1D mass-spring model was presented [38]. This Burridge-Knopoff model has been used primarily to describe the collective behavior and statistical features of EQs. It reproduces the major empirical laws of observed seismicity, i.e., earthquake recurrence, the Gutenberg-Richter law, foreshock and aftershock activities, e.g. [39-44]. In general, however, the Burridge-Knopoff model is not adequate to describe the dynamics of an individual event.

A different methodology developed by Tse and Rice [45] is based on what is known as the rate-state friction law discovered in laboratory experiments by Dieterich [46] and Ruina [47]. The main advantages of this widely used approach are (1) its simplicity (the system of two basic equations describing the rate-state frictional law can be solved analytically in the 1D approximation and is comparatively easy to analyze computationally in the 2D approximation) and (2) its universality (the model can be used to describe both EQs, e.g. [48-51], and SSEs [52-58]). However, the model can only provide a simplistic picture of rupture dynamics, which are solely determined by the ratio of the rate-state constants $a$ and $b$. Moreover, the model predicts that seismic events occur only for $b > a$, which is the regime of velocity-weakening frictional sliding. Yet, there are indications, both experimental [59] and theoretical [8, 60, 61], that frictional sliding instabilities are not confined only to velocity-weakening rocks.

We developed a model that is also suitable for the description of both EQs and SSEs [1-8]. The advantages of this model are (1) it is an intrinsically nonlinear, dynamical model, rooted in the Newtonian equations of motions; (2) all model parameters have explicit and unambiguous physical correlates; (3) it describes processes over a wide range of conditions, from very fast processes such as regular EQs down to very slow processes such as silent EQs with a few month rise time; (4) there are analytical solutions to the model for certain initial conditions; (5) the model is not computationally intensive.

The first application of our model was to a general description of the dynamics of crustal faults [1]. This approach provided quantitative values for the dynamic variables describing plate boundary EQs as well as slow earthquakes and afterslip and explained why temporal and spatial distributions of aftershocks obey, respectively, the Omori and a 1/r laws. The model was then applied to 'episodic tremor and slip' (ETS) phenomenon and successfully described the basic morphological features of ETSs [2, 4, 5]. The next application was devoted to the description of the transition process from static to dynamic frictional regimes [3]. The significant result of that article was the demonstration that the rupture velocity is defined by the shear to normal stress ratio, in quantitative argument with laboratory experiments [62]. In subsequent articles [6, 7], the governing equations were modified to include a simple velocity-dependent frictional term. Counterintuitively, the model showed that the rupture velocity does not depend on friction. Recently, we re-examined the necessary and sufficient conditions for instability of frictional sliding in the framework of our model, incorporating the rate-and-state frictional law [8]. The analysis used the standard technique of linearizing a set of equations derived from the model and investigating their behavior under small perturbations. We showed that sliding instability can develop in velocity-strengthening as well as in velocity-weakening frictional sliding, in contrast to the prevailing view that only velocity-weakening sliding leads to instability. The onset of instability is a separate topic, however. Here, we are interested in seismic dynamics assuming an instability has already developed, i.e., after nucleation of the event.

The model, to date, enables one to simulate spatial-temporal distributions of stress, slip, slip velocity, and rupture propagation velocity. There are multiple evidences, both experimental [62-64] and theoretical [65-70], suggesting that stress heterogeneity plays a crucial role in nucleation and dynamics of the rupture. The spatial distribution and magnitude of heterogeneity are the main factors that determine the dynamics of a seismic event. A distinguishing feature of our model is that the stress heterogeneity is introduced as an initial condition. We have used only a simple step-like distribution previously.

In the present study, we consider three different initial conditions that represent realistic stress heterogeneities for transform and subduction faults. Moreover, rather than using our previously simple velocity-dependent frictional term, our model now incorporates the rate-state [46, 47] and flash heating [71-73] frictional laws. We obtain the conditions that determine whether a seismic event becomes an EQ or a SSE by calculating the dependence of rupture characteristics (propagation velocity, slip velocity, displacement, and stress drop) on the stress spatial distribution and the coefficients in the frictional law used. In addition, the discovery of self-healing slip pulses [74] motivates a search for the mechanisms and/or conditions generating a pulse-like mode rather than a crack-like mode [44, 56, 75-86]. We determine such conditions in the framework of our model.

In this paper, we answer the following questions:
- What is the difference between EQs and SSEs from the point-of-view of their physical mechanisms?

- What are the critical parameters and their values which determine the precise spatial and temporal distribution of rupture velocity, slip velocity, slip, and stress drop?
- What is the role, both qualitative and quantitative, of stress heterogeneity in EQ and SSE development?
- What conditions (quantitative) determine the appearance of pulse-like ruptures versus crack-like ruptures during EQs and SSEs?
- What is the role of the rate-state frictional law in the evolution of seismic events?

Section II describes the model and the system of governing equations. Section III presents the results of simulations for the three initial conditions considered. The results are discussed in the context of known conventional results. The final section summarizes our findings.

## II.   Model

Our approach to modeling EQs and SSEs derives from the Frenkel-Kontorova (FK) model [87], as described in our previous publications [1, 3, 6, 7, 88]. Here, we briefly recap the derivation and introduce extensions to the model that are the topic of this article. We first describe the FK model that is used as the foundation. This is followed in section B by modifications we made to the FK model to adapt it to the study of frictional sliding. The sine-Gordon (sG) equation which naturally emerges is relevant to the description of slip dynamics at such a small scale as to be not only unobservable, but unnecessary for understanding the macroscopic phenomena of interest. Section C therefore outlines our derivation of sG modulation equations that are more suitable and more relevant to the description of large-scale phenomena. Sections D and E introduce the new elements incorporated in the current model. Table 1 provides values of the parameters used in the simulations.

### A. Frenkel-Kontorova model

The FK model was proposed to explain why plasticity in crystalline materials is initiated by stress that is a few orders of magnitude smaller than was expected from prevailing analyses. It assumes the presence of defects in the crystal lattice. The existence of such defects, linear dislocations, was indeed confirmed later [89]. The reason that dislocations dramatically reduce the shear stress threshold for the onset of plasticity is that the movement of a defect requires much less shear stress than the stress necessary to overcome the potential energy barrier between an atom and the lattice. The movement of a dislocation is accompanied by a local relative shift of the crystal parts. Thus, the plastic deformation occurs due to movement of multiple dislocations. The analogy would be to caterpillar-like motion. This process may be described by a nonlinear equation [89]:

$$M \frac{\partial^2 u_i}{\partial t^2} - K_b(u_{i+1} - 2u_i + u_{i-1}) + F_d \sin \frac{2\pi}{b_0} u_i = F_i - f_i, \qquad (1)$$

where $u_i$ is the shift of atom $i$ relative to its equilibrium position in the direction, $x$, of plastic deformation, $t$ is time, $b_0$ is a distance between atoms, $M$ is the mass of the atom, $F_d$ is the amplitude of the periodic force occurring due to the presence of atomic layers beneath and above the plane of plastic deformation, $F_i$ is the external force, and $f_i$ is the frictional (dissipative) force. The atoms interact with the nearest neighbors on either side via harmonic spring forces of constant $K_b$. Assuming that the crystal material has a volume density $\rho$ and an interatomic distance $b_0$ in all directions, we can express the coefficients in Eq. (1) as (see [89] for details):

$M = \rho b_0^3$, $K_b = \frac{2Gb_0}{1-v}$, $F_d = \frac{Gb_0^2}{2\pi}$, where $G$ is the shear modulus and $v$ is the Poisson ratio.

Equation (1) can then be presented in the form

$$\frac{\partial^2 (2\pi u/b_0)}{\partial (tc/b_0)^2} - \frac{\partial^2 (2\pi u/b_0)}{\partial (x/b_0)^2} + \frac{1-v}{2} \sin \frac{2\pi}{b_0} u = (\sigma_s - \sigma_f) \frac{2\pi(1-v)}{2G}, \qquad (2)$$

where the second term on the left hand side is obtained in the continuum limit approximation, $\sigma_s$ is shear stress, $\sigma_f$ is frictional force per unit area, and $c^2 = \frac{2G}{\rho(1-v)} \equiv \frac{c_l^2(1-2v)}{(1-v)^2}$, given in terms of the longitudinal acoustic velocity, $c_l$. Note that $c$ is bounded by $c_l$ and the shear wave velocity, $c_s$.

## B. FK modified for frictional slip dynamics

We apply the FK model to describe slip dynamics by modeling macroscopic friction as the result of interactions between asperities on the surfaces of two sliding solids. We consider the asperities on one of the frictional surfaces as forming a linear chain of equally spaced balls. The simplest model assumes the asperities on the opposite frictional surface form a rigid fixed substrate with the spacing between asperities equal to the ball spacing. The variable $u$ then represents the displacement of a given ball at position $x$ and time $t$ relative to its non-displaced position defined as being in the middle between two of the opposite asperities.

The maximum force between surfaces in the case of plasticity, $F_d$, becomes $(1-\nu)/2$ in Eq. 2. For modeling frictional sliding, the value of this coefficient should be reduced by the ratio between nominal and actual contact areas, which has been shown [90, 91] to be equal to the ratio $\Sigma_N/\sigma_p$, where $\Sigma_N$ is the effective normal stress (i.e., normal stress minus pore pressure) and $\sigma_p$ is the penetration hardness. The hardness of the polycrystalline materials and shear modulus are related by a simple formula, $\sigma_p \approx G/(2\pi)$ [92], hence the ratio $\dfrac{\Sigma_N}{\sigma_p} \equiv \dfrac{2\pi\Sigma_N}{G}$. We also have $\sigma_f = \mu\Sigma_N$ in terms of the standard coefficient of friction, $\mu$. Then, to describe frictional sliding, Eq. (2) becomes

$$\frac{\partial^2(2\pi u/b_0)}{\partial(tc/b_0)^2} - \frac{\partial^2(2\pi u/b_0)}{\partial(x/b_0)^2} + \frac{2\pi\Sigma_N}{G}\frac{1-\nu}{2}\sin\frac{2\pi}{b_0}u = (\sigma_s - \sigma_f)\frac{2\pi(1-\nu)}{2G}, \tag{3}$$

where now $b_0$ is the typical distance between asperities. Dividing both sides by $\dfrac{2\pi\Sigma_N}{G}\dfrac{1-\nu}{2}$ and introducing dimensionless variables $\tilde{u}$ in units of $b_0/(2\pi)$, $\tilde{x}$ in units of $[\dfrac{2G}{2\pi\Sigma_N(1-\nu)}]^{1/2}b_0$, $\tilde{t}$ in units of $[\dfrac{2G}{2\pi\Sigma_N(1-\nu)}]^{1/2}\dfrac{b_0}{c}$, and $\tilde{\sigma}_S$ and $\tilde{\sigma}_f = \mu$ in units of $\Sigma_N$, we obtain the well-known sine-Gordon (sG) equation:

$$\frac{\partial^2\tilde{u}}{\partial\tilde{t}^2} - \frac{\partial^2\tilde{u}}{\partial\tilde{x}^2} + \sin\tilde{u} = \tilde{\sigma}_s - \tilde{\sigma}_f, \tag{4}$$

## C. The sine-Gordon modulation equation

There are three fundamental solutions of the sG equation: kinks (or solitons), plasma waves, and breathers. In the framework of plasticity, a soliton represents a linear dislocation. In the context of friction, we model a soliton as an area of accumulated, enhanced stress nucleated on the surface by shear stress applied to an inhomogeneous surface. We will use the term

"macroscopic dislocation" to designate such areas. In crystals, the displacement of a dislocation under the applied shear stress (its "mobility") is much larger than the corresponding translation of the entire atomic plane. In the same way, the mobility of a macroscopic dislocation over the "bumpy road" of a frictional surface (i.e., the quasi-periodic substrate) is larger than the mobility of the whole surface. In both cases, the displacement of a dislocation (a pre-stressed area) requires much less external stress. Therefore, in our model, the relative sliding of two bodies along a planar interface, which occurs due to the movement of macroscopic dislocations, can be initiated by external stress smaller than $\mu\Sigma_N$, which is consistent with laboratory experiments [62]. We note that a dislocation may propagate with any velocity ranging from zero to $c$.

The sG equation has been widely used to describe nanoscale friction (see, e.g., [93] and references therein). It has also been applied to macroscopic friction [94-96], but it is not particularly useful for this case. The sG equation is relevant for quantitatively describing the dynamics of only a few dislocations. It is not useful for describing plastic deformation in a macroscopic crystal, which may include millions of dislocations per centimeter. The same is true for macroscopic friction. If one requires a detailed description of the relative movement of two bodies at a scale of a few asperities, the sG equation is the right tool. However, for objects which include millions of asperities per cm$^2$ (i.e., thousands of macroscopic dislocations), the pure sG is not an appropriate instrument. The details it provides of dynamics at the microscopic level are not directly observable in standard macroscopic experiments. The actual position of any particular dislocation is not important.

The dynamics of a sequence of a large number of dislocations are better described by variables averaged in space and time over the dislocation periods. Witham developed a technique for transforming an equation describing the behavior of given variables to a system of equations describing the behavior of the variables averaged over space and time [97]. Applying this method to the sG equation, we derived a system of sG modulation equations [88], expressed again in dimensionless variables, which is suitable for describing macroscopic friction:

$$-\frac{4}{\pi}\frac{K}{[m(1-U^2)]^{1/2}}[U_t\frac{\eta}{U^2-1}+m_t\frac{U}{2m}+U_x\frac{U\eta}{U^2-1}+m_x\frac{1}{2m}]=\Sigma_S-\mu$$
$$U_t\frac{U}{U^2-1}+m_t\frac{\eta}{2mm_1}+U_x\frac{1}{U^2-1}+m_x\frac{U\eta}{2mm_1}=0 \quad (5)$$

Here $m$ is the modulus of the Jacoby elliptic function ($0 < m \leq 1$), $m_1 = 1 - m$, $\eta = E/K$, where $K(m)$ $K(m)$ and $E(m)$ $E(m)$ are the complete elliptic integrals of the first and second kind, respectively, $U$ is the wave velocity in units of velocity $c$, with $-1 \leq U \leq 1$. Subscripts $x$ and $t$ denote the derivative with respect to the associated variable. The friction coefficient $\mu$ and (dimensionless) shear stress $\Sigma_s$ now represent the more relevant macroscopic averages over the dislocation periods.

### D. Relation to macroscopic physical variables

The variables $m(x,t)$ and $U(x,t)$ obtained by solving Eqs. (5) can be associated with macroscopic physical variables of interest. The (dimensionless) shear stress and slip velocity $W$ (in units of $c(\frac{\Sigma_N}{2\pi G}\frac{1-\nu}{2})^{1/2}$ are given by [1]

$$\Sigma_S = \frac{\pi}{K[m(1-U^2)]^{1/2}}, \tag{6}$$

$$W = U\Sigma_S. \tag{7}$$

Slip $S$, which is directly related to the suitable average for $\tilde{u}(x,t)$ in Eq. (4) that motivated the derivation of Eqs. (5), is simply

$$S = \int_0^t W dt, \tag{8}$$

while stress drop, $\Delta\Sigma$, is

$$\Delta\Sigma(x,t) = \Sigma_S(x,t) - \Sigma_S(x,0). \tag{9}$$

Given initial values at $t = 0$ for the spatial distribution of $\Sigma_S$ and $W$, one can use Eqs. (6) and (7) to obtain initial values for $m(x,0)$ and $U(x,0)$. One must also input an initial value for friction, which is the topic of the next section. Equations (5) can then be written using only the two variables $m$ and $U$ and solved numerically. Equations (6) and (7) then give the redistribution of shear stress, $\Sigma_S(x,t)$ and slip velocity, $W(x,t)$, from which are derived the slip and stress drop using Eqs. (8) and (9). Any of these spatial and temporal profiles will show the evolving rupture front or edge as a clear discontinuity delineating the modified values in the rupture at times $t$ from the initial values at positions ahead of the front. Then the average rupture velocity, $V$, is simply the distance the rupture front has travelled in the corresponding time interval.

**E. Friction**

In previous work, our model has either not included friction or simply set it proportional to slip velocity [6, 7]. In the present work, we incorporate a more physically realistic friction into our model. For slip velocity $W \leq 0.1$ m/s, we will use the well-known Dieterich-Ruina rate-state model [46, 47] for the coefficient of friction, $\mu$:

$$\mu(W,\theta) = \mu_0 + a\ln(\frac{W}{W_0}) + b\ln(\frac{W_0\theta}{D_c})$$
$$\frac{d\theta}{dt} = 1 - \frac{W\theta}{D_c}, \tag{10}$$

where $a$ and $b$ are empirical dimensionless constants, $\theta$ is the state variable in units of $b_0 G/(2\pi \Sigma_N c)$, and $D_c$ is the characteristic slip distance. The connection between parameters $b_0$ and $D_c$ is unknown. We assume that $b_0/(2\pi D_c) \approx 1$. The latter is true if asperities have a sinusoidal shape and if $D_c$ is equal to the root-mean-square roughness of the surface. When $\dot{\theta} = 0$, we have the steady state values $\theta_{ss} = D_c/W$, in which case, the steady state friction $\mu_{ss}$ is

$$\mu(W,\theta_{ss}) \equiv \mu_{ss}(W) = \mu_0 + (a-b)\ln(\frac{W}{W_0}),$$

giving $\mu_{ss}(W_0) = \mu_0$. Representative reference values for $\mu_0$ and $W_0$ are provided in Table 1.

For $W > 0.1$ m/s, flash heating leads to a low frictional strength [69-71]. To describe friction for slip velocities $W > W_{flash} = 0.1$ m/s, Rice [98] proposed a simple formula:

$$\mu(W) = [\mu_{ss}(W = W_{flash}) - \mu_{flash}]\frac{W_{flash}}{W} + \mu_{flash},$$

where $\mu_{flash}$ is the minimum value of the friction coefficient resulting from flash heating, determined from experiments to be ~0.2 [71-73]. However, the effective friction also tends to increase due to radiation damping (related to the power of the radiated seismic waves) as the velocity $W$ increases. To include this competing effect, we modify the previous relation to

$$\mu(W) = [(\mu_{ss}(W_{flash}) - \mu_{flash})\frac{W_{flash}}{W} + \mu_{flash}][1 + \alpha(\frac{W - W_{flash}}{W_{flash}})^2], \tag{11}$$

where  $\alpha = \dfrac{(W_{flash}/W_m)^2(\mu_0 - \mu_{flash})}{(1-W_{flash}/W_m)[(\mu_0 - \mu_{flash})W_{flash}/W_m + \mu_{flash}]}$ ,  and (dimensionless) $W_m = 1.1 \times 10^{-2}$

(equal to 1 m/s in physical units) is a representative average value for the maximum slip velocity within a rupture for an EQ event. This expression for $\alpha$ is derived to satisfy the condition $\mu(W_m) = \mu_{ss}(W_{flash})$. An example of the dependence of friction on slip velocity, $\mu(W)$, is shown in Fig. 1 for the particular value $\theta = \theta_{ss}$ (steady state). For the velocity weakening case chosen ($a < b$, see Table 1), friction steadily decreases, as given by Eqs. (10), for slip velocities $W \leq W_{flash}$. Then as $W$ increases to $W > W_{flash}$, friction abruptly decreases further and finally sharply increases with further increase of $W$, as given by Eq. (11).

## F. Initial conditions used in the model

To solve the system of Eqs. (5) – (11), two initial conditions must be specified. The initial slip velocity is chosen to be the average slip velocity along a fault, $W_0 = W(x, t = 0)$. A typical value is $W_0$ = 3 cm/yr. In contrast to conventional approaches, we will use spatially heterogeneous shear stress as an initial condition. The reason for this is that sliding instabilities develop in locations with a gradient in the ratio of shear to effective normal stress [8]. Hereafter we will use the term "stress ratio" instead of "shear to effective normal stress ratio".

Three configurations of the initial stress distribution will be used for the simulations (shear stress applied, without loss of generality, in the positive *x*-direction):

$$\Sigma_S(x, t = 0) = \Sigma_{S0} - \dfrac{\delta\Sigma}{\pi}\tan^{-1}(\beta x) \tag{12a}$$

$$\Sigma_S(x, t = 0) = \Sigma_{S0} - \dfrac{\delta\Sigma_b}{\pi}\tan^{-1}(\beta x)e^{-\zeta x} \tag{12b}$$

$$\Sigma_S(x, t = 0) = \Sigma_{S0} + \dfrac{\delta\Sigma}{2\cosh(\zeta x)}, \tag{12c}$$

where $\Sigma_{S0}$ is the value of the average shear stress in the distribution, $\delta\Sigma$ is the difference between the maximum and minimum stress, $\beta$ regulates the slope of the profile connecting the maximum and minimum stress values, and $\zeta$ determines the spatial extent of variations in the stress distribution. In order to properly compare the simulation results, the maximum stress in all three cases is set equal, which requires $\delta\Sigma_b = \frac{\pi}{2} \frac{e^{-\zeta|x_{max}|}}{\tan^{-1}(\beta|x_{max}|)} \delta\Sigma$ in Eq. (12b), where $x_{max}$ is the location of the maximum in the stress profile.

Spatial profiles for the three initial conditions are illustrated in Figs. 2(a-c). The first two configurations mimic the spatial distribution of shear stress around an obstacle, which prevents relative sliding of the plates along a fault. Stress accumulates on one side of the obstacle and there is a stress deficit on the other side. The third configuration approximates effects due to localized reductions in the effective normal stress, due, for example, to changes in pore pressure or "nonplanar interfaces being compressed into full contact" (e.g. [99]).

A FORTRAN code was developed to obtain numerical solutions. The code allows one to study the evolution of frictional processes employing the predictor-corrector scheme. Typical values for the effective normal stress used in the model are 60 MPa for regular EQs and 100 MPa for SSEs, based on the lithostatic pressure at depth $h$, equal to $\rho g h$.

### III. Results and Discussion

New results from the updated model are presented here. In Section A, we show that the model is capable of simulating a wide range of seismic events, from a typical regular EQ with slip velocity ~ 1 m/s and rupture velocity ~ $c_s$ to a SSE with slip velocity ~ 1 m/year and rupture velocity ~1 km/day. Then we determine the specific conditions (initial stress spatial distribution and frictional parameters) that produce either an EQ or SSE. In Section B, we provide examples to show the model can account for both crack-like and pulse-like ruptures. We then identify the conditions that control the rupture mode. In Section C, we consider additional differences in rupture shape that result from differences in the initial stress heterogeneity.

### A. Conditions for regular EQ versus SSE

Let us first consider a particularly simple example of stress heterogeneity in which the initial stress is the non-localized step-like function given by Eq. 12(a) and depicted in Fig. 2(a). We assume that a rupture (seismic event) is initiated at $t=0$ and at the position $x=0$ where the gradient in the stress ratio is maximum.

Values $\Sigma_{S0} = 0.7$ and $\delta\Sigma = 0.1$ give results that are typical for a regular EQ. Figure 3 shows the resulting solutions for the spatial distribution of shear stress, $\Sigma_S$, and slip velocity, $W$, at three different times. Further details of the spatial-temporal distribution of $\Sigma_S$ and $W$, as well as stress drop, $\delta\Sigma$, and slip, $S$, are shown in Fig. 4. Recall that the average rupture velocity, $V$, is the distance the front travels (the location of the rupture edge, as shown in the figures) divided by the time travelled. As one can see, the rupture propagates in two directions with slightly different velocities that we calculate as $V^- = 3.30$ km/s $= 0.9c_s$ to the left and $V^+ = 2.24$ km/s $= 0.61c_s$ to the right. The rupture propagates faster in the region of greater shear stress, $V^- > V^+$, consistent with our previous finding [1, 3]. As can be seen from Fig. 4, $W$ quickly reaches a value of ~1m/s near the initiation point $x = 0$ and remains constant at this point thereafter. In the rest of the rupture area, $W$ is ~0.8 m/s. The slip is maximum at the initiation point and reaches ~10 m after ~10 sec. The maximum stress drop shown in the figure is about 3 MPa, which is about 7% of the average initial shear stress of ~42 MPa.

Performing the same calculations using the values $\Sigma_{S0} = 0.18$ and $\delta\Sigma = 5\times10^{-4}$ produces a SSE. The results, converted to physical units using the SSE normalization factor $\Sigma_N = 100$ MPa, are shown in Fig. 5. The slip velocity $W \approx 2.3$ m/year, and the rupture velocities are $V^- \approx V^+ \approx 0.88$ km/day, which are, respectively, seven and five orders of magnitude less than slip and rupture velocities of a regular EQ. The stress drop is about 40 KPa, two orders of magnitude less than for a regular EQ. The stress drop could be even smaller than given by our example using $\Sigma_N = 100$ MPa, since there are indications [4, 5] that the effective normal stress in subduction zones may be much less than the lithostatic pressure.

Thus, the differences between the examples illustrated in Figs. 4 and 5 indicate that the dynamic variables describing the rupture depend critically on the values of $\Sigma_{S0}$ and $\delta\Sigma$ relative to $\Sigma_N$. To examine this dependence in detail, we solved Eqs. (5 – 12) for wide ranges of $\Sigma_{S0}$ (from 0.15 to 0.9) and $\delta\Sigma$ (from 0.001 to 0.1). The initial stress distribution of Eq. 12a, Fig. 2(a) and friction, Eqs. (10) and (11), were inputs to the model. They were calculated using parameters listed in Table 1, which gives the velocity-weakening ($a < b$) profile for friction (see Fig. 1). Results are plotted in Fig. 6. The maximum calculated slip velocity in the rupture, $W_{max}$, is plotted in the upper panel for different values of $\delta\Sigma$ as a function of the maximum stress $\Sigma_{max} = \Sigma_{S0} + \delta\Sigma/2$, normalized to $\Sigma_N$. The solid curves are the results of the friction model and show the effects of changing from rate-state, Eq. (10) to flash heating, Eq. (11), for values of $\Sigma_{max}$ that give $W_{max} \geq W_{flash}$. The dashed curves are the results of using only rate-state friction for all $\Sigma_{max}$. The rupture velocity $V^-$ of the front propagating in the direction of decreasing $x$ is plotted in the lower panel.

Changing $\Sigma_{max}/\Sigma_N$ by less than an order of magnitude changes $W_{max}$ by about nine orders of magnitude and $V^-$ by about seven orders. When $W_{max} = W_{flash}$, friction abruptly decreases due to flash heating, and the (dimensionless) slip velocity jumps to values $W_{max} \sim 10^{-2}$, (see abrupt steps in the solid curves of the upper panel) corresponding to physical values $\geq 1$ m/s that are characteristic of EQs. There is also a jump in $V^-$ for the same transition values of $\Sigma_{max}/\Sigma_N$, as shown in the lower panel. If flash heating is not considered and only rate-state friction is used for all $W$, the slip and rupture velocities are much smaller with increasing $\Sigma_{max}/\Sigma_N$ (see dotted curves in the upper panel and curves in the bottom right rectangle of the lower panel). Almost identical results (not shown) are obtained for the other two initial stress profiles, Eqs. 12*b* and *c*. These results do not appear to depend sensitively on the exact profile of the shear stress.

Thus, we conclude that in our model, the type of seismic event, i.e. EQ or SSE, is determined primarily by the stress ratio $\Sigma_{max}/\Sigma_N$. Furthermore, EQs, i.e., slip with maximum slip velocity in the rupture $W_{max} \sim 1$ m/s (~0.01 in dimensionless units), only occur if there is a dramatic jump in slip velocity due to a pronounced decrease in friction due to flash heating or other mechanisms when the velocity reaches $W_{flash}$. Rate-state friction alone does not provide the necessary slip velocity using experimental values for the parameters *a* and *b* unless unrealistically large values $\Sigma_{max} \geq 0.9\Sigma_N$ are used.

In previous work, we showed that, in contrast to common assumption, sliding instability can develop for velocity-strengthening (*a* > *b*), as well as velocity-weakening (*a* < *b*), frictional sliding [8]. Instability in velocity-strengthening sliding was found to be the result of a sufficiently large gradient in the stress ratio. Velocity-weakening instability is, nonetheless, more probable since nucleation lengths are smaller than for velocity-strengthening materials. Here, for $W < W_{flash}$, we find in addition that when the rupture is already nucleated, i.e., an instability has developed, the dynamics of the rupture are the same for the velocity-weakening and velocity-strengthening cases. In other words, the dynamics of the rupture do not depend on the relative values of *a* and *b* in the rate-state law. The foregoing holds for all SSEs since $W < W_{flash}$ in all these seismic events.

### B. Conditions for crack-like versus self-healing pulse rupture

In the previous section, we found that the primary determinants of whether a rupture event manifests as an EQ or a SSE are the following values used as inputs to the model for the initial stress distribution: the maximum value in shear to normal stress ratio, $\Sigma_{max}/\Sigma_N$, and $\delta\Sigma$, the difference between the maximum and minimum stress. The results were not sensitive to other details of the stress profile, so we focused on examples obtained using the initial stress distribution of Eq. 12*a*. Having established that the model can account for both EQs and SSEs, as well as providing the conditions that distinguish between them (e.g., Fig. 6), we now look for rupture properties that *do* depend on particulars of the spatial heterogeneity in the initial stress profile.

Note that all rupture examples considered so far have been crack-like, i.e., the slip velocity at any point within the rupture is approximately constant in time. In the context of a seismic event it means that the rise time equals the slip time at a rupture initiation point. Heaton [74] showed that for some EQs the slip time at some point of a rupture is much less than the EQ rise time indicating that at least in these particular areas the rupture takes the form of a self-healing pulse.

Consider the spatially localized inhomogeneity in the initial stress given by Eq. 12*b* (see also Fig. 2*b*). The results of simulations using values ($\Sigma_{S0} = 0.7$, $\delta\Sigma = 0.17$) that give an EQ are shown in Figs. 7 and 8. We find a crack-like rupture that propagates at velocities $V^{\pm}$ in opposite directions. The localized slip to the right, i.e. in the shear stress direction, causes a stress drop at the left side of the rupture and stress rise at the right side (Fig. 7 illustrates the development of this structure). As a result, the disturbance area expands in both directions. The maximum slip velocity in the rupture is about 1 m/s, as expected for an EQ, and the maximum stress drop is about 4 MPa.

The results are similar to those obtained using the initial condition of Eq. 12*a* (see Figs. 3 and 4) both qualitatively and quantitatively, i.e. the rupture is an expanding crack. Investigating further, we find the rupture is always crack-like for the initial stress profile of Eq. 12*a*, which is asymmetrically homogeneous outside the heterogeneous region. For Eq. 12*b*, the rupture is crack-like only if $\Sigma_{S0}$ and $\delta\Sigma$ are large enough that the system evolves to the abrupt velocity-weakening (flash heating) conditions, so that the maximum slip velocity in the rupture becomes larger than $W_{flash} = 0.1$ m/s. For smaller values of $\Sigma_{S0}$ and/or $\delta\Sigma$ the system can behave very differently.

We consider the previous example, but reduce $\delta\Sigma$ by a factor of 10 to produce an SSE, with results shown in Figs. 9 and 10. The rupture now expands in both directions by Heaton-like (self-healing) pulses, i.e. the slip velocity at the rupture fronts is much larger than inside the rupture area. The maximum slip velocity is about 0.06 m/s, and the system never switches from rate-state to flash heating. The maximum slip and the maximum stress drop are centered at the initiation point, $x = 0$. The rupture propagation velocity is a factor 1.3 smaller than in the previously discussed case of larger maximum slip velocity (Fig. 8), where the rupture is crack-like.

Further analysis shows, in addition to $W < W_{flash}$, a pulse-like rupture requires a spatially localized heterogeneity in the initial stress profile. The solution should be obtained over a region that exceeds the extent of the heterogeneity by at least a factor of ~3 in order to distinguish the pulse-like character of the rupture. If flash heating is included in the simulations and $\Sigma_{S0}$ and $\delta\Sigma$ are large enough to drive the system to the conditions where $W > W_{flash}$, then the rupture is crack-like. However, if flash heating is not included in the model, the rupture is always pulse-like. The abrupt reduction of friction due to flash heating prevents the development of a pulse-like rupture. The conditions that distinguish between the occurrence of crack-like and pulse-like ruptures were obtained from simulations and are summarized in Fig. 11.

### C. Nuances in rupture characteristics due to stress heterogeneity

We have found the conditions that determine whether a seismic event is an EQ or SSE (see Fig. 6) and whether the resulting rupture is crack-like or pulse-like (Fig. 11). In particular, we found that these results were not sensitive to details of the stress profile, depending only on $\Sigma_{S0}$ and $\delta\Sigma$. In this section, we investigate effects due to different types of initial stress profile.

Consider the spatially localized inhomogeneity in the initial stress given by Eq. 12$c$ (see also Fig. 2$c$). The results of simulations using values $\Sigma_{S0} = 0.7$ and $\delta\Sigma = 0.05$ that give an EQ are shown in Fig. 12. As expected for these parameters, the rupture is crack-like. However, in contrast to the crack-like ruptures obtained previously, the region of maximum slip and maximum stress drop is no longer centered at the rupture initiation point, $x = 0$, but shifts to the right.

For values ($\Sigma_{S0} = 0.2$, $\delta\Sigma = 5 \times 10^{-4}$) that give a SSE (convert to physical units using $\Sigma_N = 100$ MPa), the rupture, shown in Fig. 13, is pulse-like, as expected. However, only one slip pulse is formed. There is no slip for points $x < 0$ even though there are still two stress pulses, one travelling to the left (opposite to the shear stress direction) and one to the right (in the direction of shear stress).

The differences in both crack-like and pulse-like behavior demonstrated using different initial stress profiles provide information on the processes underlying a seismic event. As noted, the initial conditions of Eqs. 12a and 12b mimic the spatial distribution of shear stress around an obstacle, which prevents relative sliding of the plates along a fault. The initial condition of Eq. 12c reflects localized reductions in the effective normal stress

Our model also provides a quantitative description of episodic tremor and slip (ETS) [2, 4, 5], defined as the slow propagation of a single slip pulse along a subduction fault accompanied by massive bursts of nonvolcanic tremor. This periodic phenomenon has been observed in virtually all major subduction zones [16, 100-104]. In the Cascadia area, the pulse slip is 3-5 cm and pulse propagation velocity is ~ 10 km/day [105]. In the current work, we have provided a possible mechanism for such pulses, which can be produced by choosing appropriate values of $\Sigma_S$ and $\delta\Sigma$, as shown in Fig. 13.

### IV. Conclusion

The inherently nonlinear model presented here describes the spatial-temporal distribution of stress, slip velocity, slip, and rupture front propagation velocity resulting from input values of physical variables taken from experiment relevant to the study of earthquakes. There are no adjustable parameters in the model, and it is able to describe the dynamics of both regular EQs and SSEs, including ETSs. In our previous work, frictional effects were modeled using a simple velocity-dependent term. The initial stress distribution was a simple step-like function. Here, we provided a more sophisticated treatment of friction and focused on the importance of stress heterogeneity in rupture nucleation and dynamics, as demonstrated in laboratory experiments [62-64] and suggested by various models [65-70]. To that end, (a) rate-state and flash-heating friction laws are now coupled with the model's system of governing equations and (b) three different heterogeneous spatial distributions for the shear to normal stress ratio have been considered as initial conditions.

The results have a fairly simple organizational structure that can be categorized according to the frictional law and the initial stress distribution. Within the context of these broad influences, our analysis shows that the type of seismic event (regular EQ or SSE), rupture mode (crack-like or pulse-like ruptures), and rupture velocity depend solely on (1) the averaged stress ratio (ratio of shear to effective normal stress); and (2) the gradient in this ratio.

In section III A, we first investigated implications of the frictional law used in the model. The abrupt decrease of friction, which occurs at slip velocities $W \geq W_{flash} = 0.1$ m/s due to flash heating [71-73] or other mechanisms [106, 107], is crucial for the occurrence of EQs. Otherwise, only SSEs are possible. One caveat is that rate-state friction alone, using experimental values for its parameters *a* and *b* (see Table 1), could produce EQs if one allows unrealistically large values of the shear to normal stress (greater than ~ 0.9). Other models [86, 108] have also noted the importance of abrupt velocity-weakening friction at large slip velocity. We then quantified the maximum slip velocity, $W_{max}$, as a function of the stress and gradient to find the values that give $W_{max} \geq W_{flash}$, resulting in EQs, or $W_{max} < W_{flash}$, resulting in SSEs (see Fig. 6).

In section III B, we next investigated the effect of spatial heterogeneity in the stress profile and found that, for localized heterogeneities as in Eqs. 12 (b) and (c), shown in Figs.2 (b) and (c), EQs are always crack-like and SSEs are always pulse-like. We then calculated the conditions resulting in a particular rupture mode, as shown in Fig. 11. These results may be applicable to more general stress distributions. The stress profiles used in the simulations were either symmetric or anti-symmetric, and a range of localized stress distributions can be constructed from linear combinations of these two shapes. The anti-symmetric step-like distribution of Eq. 12(a), which extends indefinitely in both directions and is therefore not localized, produces only crack-like ruptures. Various models [44, 56, 75-86] have also considered the conditions leading to crack-like ruptures or self-healing (pulse-like) ruptures. Our model is distinctive in linking the type of rupture mode to spatial heterogeneity in the stress profile and providing quantitative values for the initial stress ratio and its gradient that produce a given rupture mode.

Upon further investigation of rupture modes, we found, in section III C, that a symmetric localized stress distribution always results in a single-pulse rupture mode for SSEs, which is suggestive of ETSs. The heterogeneity profile also determines the position of the maximum slip for EQs, which, as shown, are always crack-like in our model. For the anti-symmetric stress profiles considered in Figs. 2(*a*) and (*b*), the location of the maximum displacement is at the rupture initiation point. For the symmetric stress profile shown in Fig. 2(*c*), the maximum displacement is shifted in the positive *x*-direction, which we have chosen as the direction of the applied shear stress.

Rupture velocity is also an important characteristic of EQs studied experimentally [109, 110] and by modeling [111-114]. We find that the rupture velocity for regular EQs ranges from a fraction of the shear wave velocity up to the supershear velocity. The range for SSEs is much wider, from $10^{-7} c_s$ to $10^{-1} c_s$. The maximum slip velocity within the rupture for regular EQs is $\geq$ 1 m/s. For SSEs, slip velocities range widely from a few cm/year up to 0.1 m/s. Thus, our modeling of rupture and slip velocities are consistent with the observational data.

The importance of obtaining a value for the stress drop during EQs and SSEs is of considerable interest, e.g. [24, 115]. Using typical parameters for EQs and SSEs, our simulations show that stress drop varies in the range from 1% to 10% of the initial shear stress for regular EQs and from 0.1% to 10% for SSEs. Generally, due to differences in the initial shear stress between EQs and SSEs, the stress drop in SSEs is one to three orders of magnitude less than in EQs, consistent with other estimates [24].

Thus, various features of seismic events predicted by the model are consistent with basic results of conventional models. However, there are also important differences. (1) As shown previously [8], in the framework of our model, the development of instability is possible not only for velocity-weakening rocks, as given by conventional models, but also for velocity-strengthening rocks. The nucleation size depends on values of the rate-state parameters *a* and *b*, as usual, but also on the value of the gradient in the stress ratio. (2) New simulations presented here show that *after* an instability develops, rupture dynamics do not depend on the parameters *a* and *b*. It does not matter whether frictional sliding is velocity-weakening (*a* < *b*) or velocity-strengthening (*a* > *b*). Popular models based on the rate-state frictional law [45, 48-58] are not consistent with these results.

In summary, the critical factor determining the spatial and temporal distribution of the rupture velocity, slip velocity, slip, and stress drop, which characterize the dynamics of EQs and SSEs, is the initial spatial distribution of the stress ratio (which includes the gradient in the ratio) prior to the onset of instability in frictional sliding. We have provided quantitative values for the distributions that determine a wide range of observed EQ and SSE phenomena. We also note that our model provides a quantitative description of certain features of ETS phenomenon (see Fig. 13 and the articles by Gershenzon et al. [2] and Gershenzon and Bambakidis [5]). However, others features of ETSs, such as rapid tremor propagation in the slip-parallel direction [105] and reverse tremor migration [116], can be described quantitatively only in 2D models, currently under development for presentation in a separate article (Part II).

## ACKNOWLEDGMENT

We thank the reviewers for elucidating limitations in the submitted manuscript.

**Tables**

Table 1. Values of the parameters used in simulations

| variable | value | variable | value |
| --- | --- | --- | --- |
| $G$ | 30 GPa | $W_{flash}$ | 0.1 m/s |
| $v$ | 0.3 | $W_m$ | 1 m/s |
| $c_l$ | 6 km/s | $W_0$ | 3 cm/year |
| $c_s$ | 3.67 km/s | $\Sigma_N$ (EQ) | 60 MPa |
| $c$ | 5.66 km/s | $\Sigma_N$ (SSE) | 100 MPa |
| $b_0$ | 3 cm | $\Sigma_{S0}$ | 0.15 – 0.9 |
| $a$ | 0.01 | $\delta\Sigma$ | 0.001– 0.1 |
| $b$ | 0.015 | $\beta$ | 0.1 |

| $\mu_{flash}$ | 0.2 | $\zeta$ | 0.01 |

# Figures

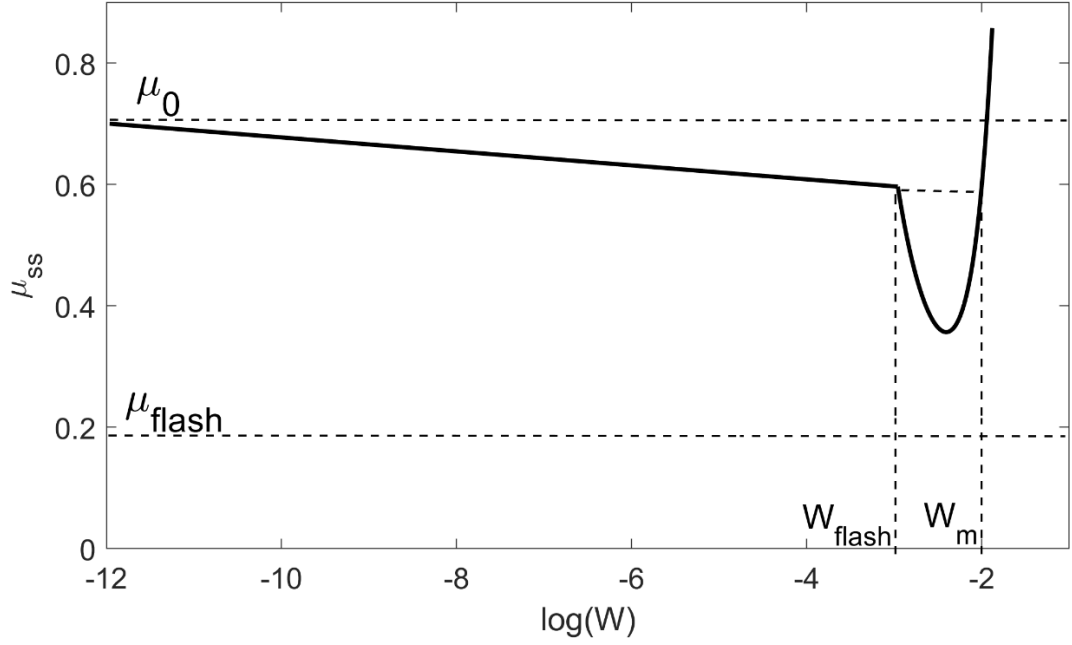

Figure 1. The steady state $\theta = \theta_{ss}$ is used as an example of the velocity-dependent friction coefficient, plotted as a function of (dimensionless) slip velocity, $W$: $\mu(W, \theta = D_c/W) \equiv \mu_{ss}(W)$ as given in Eq. (10) for $W \leq W_{flash}$ and $\mu_{ss}(W) \equiv \mu(W)$ from Eq. (11) for $W > W_{flash}$, using parameters listed in Table 1 which give a velocity-weakening ($a < b$) example for rate-state friction. The value $\mu_0 = \mu_{SS}(W_0)$, where $W_0 \approx 10^{-12}$ in dimensionless units has been chosen as a representative value for the average slip velocity along a fault, corresponding to ~3 cm/yr.

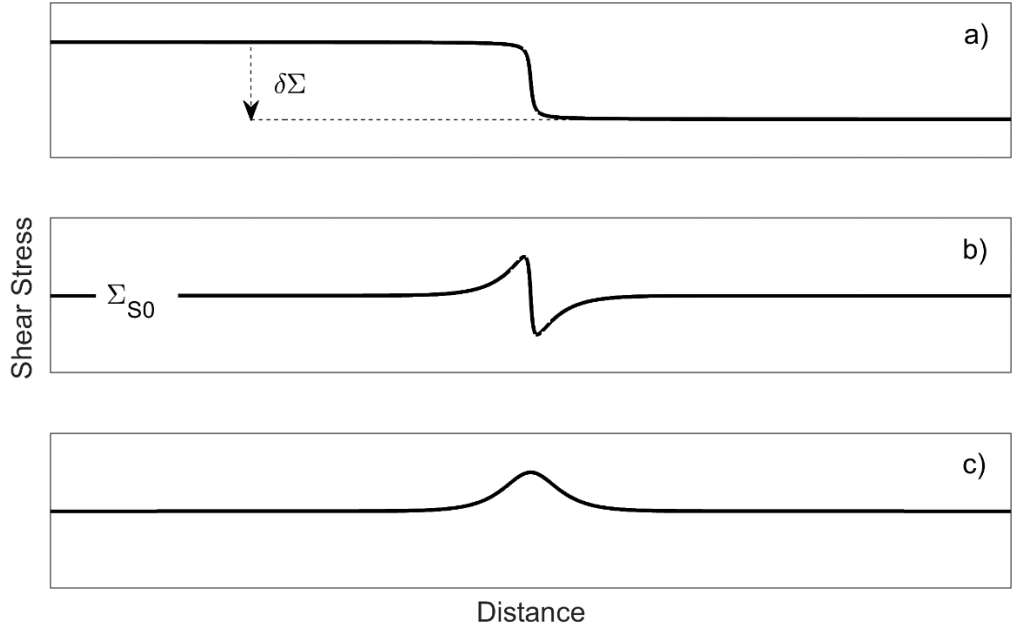

Figure 2. Spatial distribution of initial shear stress for the three configurations given in Eqs. (12a-c), applied, without loss of generality, in the positive *x*-direction. In each panel, $\delta\Sigma = \Sigma_{max} - \Sigma_{min}$, illustrated in particular for case (a), and $\Sigma_{S0}$ is the average shear stress value, illustrated in particular for case (b). The first two configurations mimic the spatial distribution of shear stress around an obstacle. The third configuration was chosen to represent effects due to a localized reduction of effective normal stress.

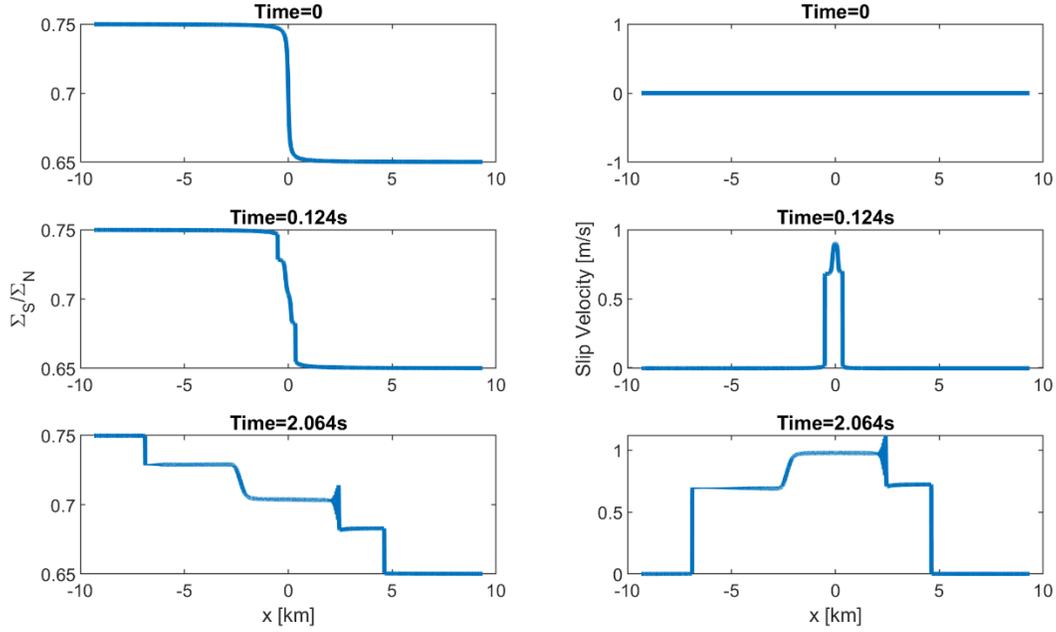

Figure 3. The spatial distribution of shear stress, $\Sigma_S$, and slip velocity, $W$, are shown at three different times, as calculated from our model, Eqs. (5-12). Solutions were obtained using the initial stress distribution of Eq. 12$a$ and values listed in Table 1. The example values $\Sigma_{s0} = 0.7$ and $\delta\Sigma = 0.1$ give results that are typical for regular EQs. The average rupture velocity, $V$, is simply the location of the rupture front (as measured from the initiation point $x = 0$) divided by the time travelled. The region of non-zero slip velocity propagates to the left at $V^- = 3.34\,\text{km/s} = 0.61c_s$, and another region of non-zero slip velocity propagates to the right at $V^+ = 2.21\,\text{km/s} = 0.40c_s$. The same holds true for regions of reduced and increased stress.

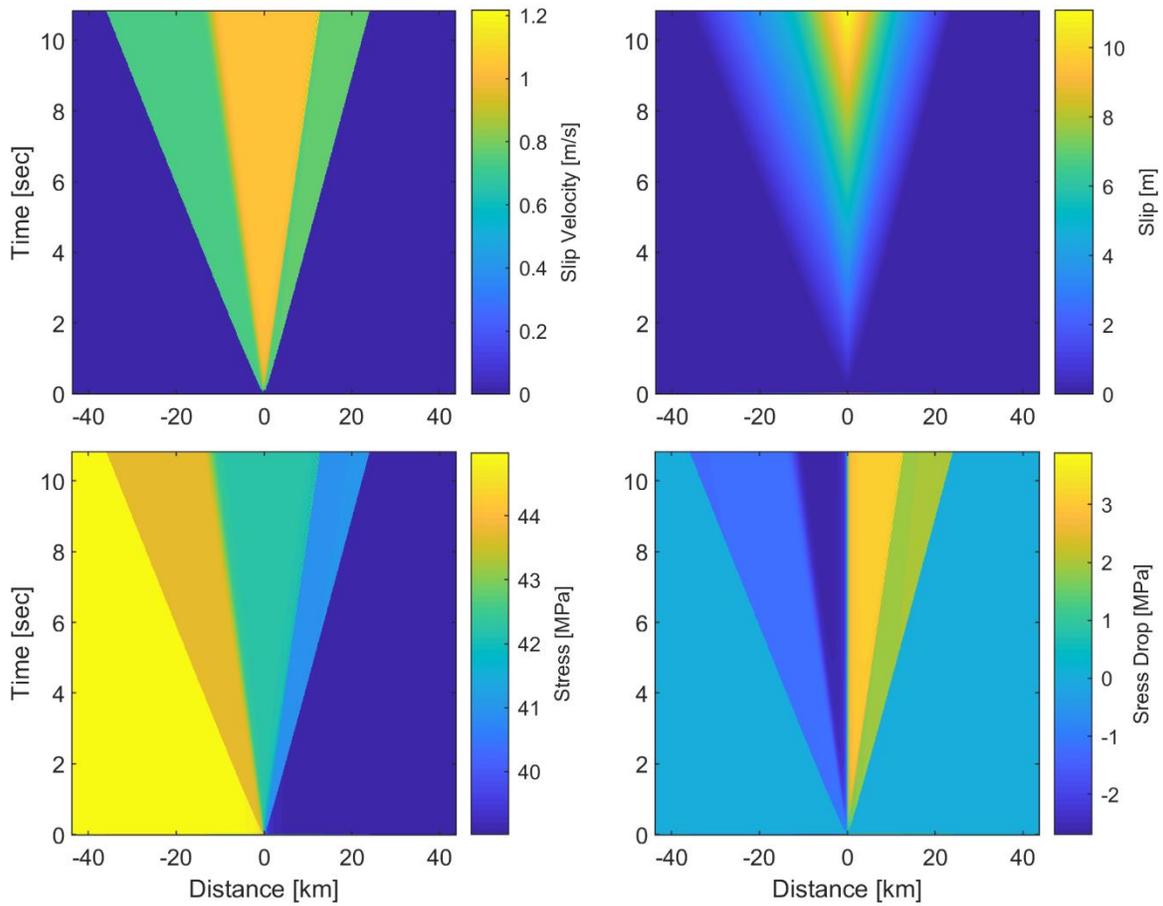

Figure 4. As in Fig. 3, showing further detail for the spatial-temporal distribution of shear stress, $\Sigma_S$, and slip velocity, $W$, as well as stress drop, $\delta\Sigma$, and slip, $S$ for a regular EQ. The rupture front is clearly evident in each panel as the discontinuity between unchanged initial values, located at positions that are beyond the rupture edge, and modified values within the rupture boundary. The front, which clearly must be the same for any of the variables shown, travels in this case to the left at velocity $V^-$ and to the right at $V^+$, as obtained in Fig. 3.

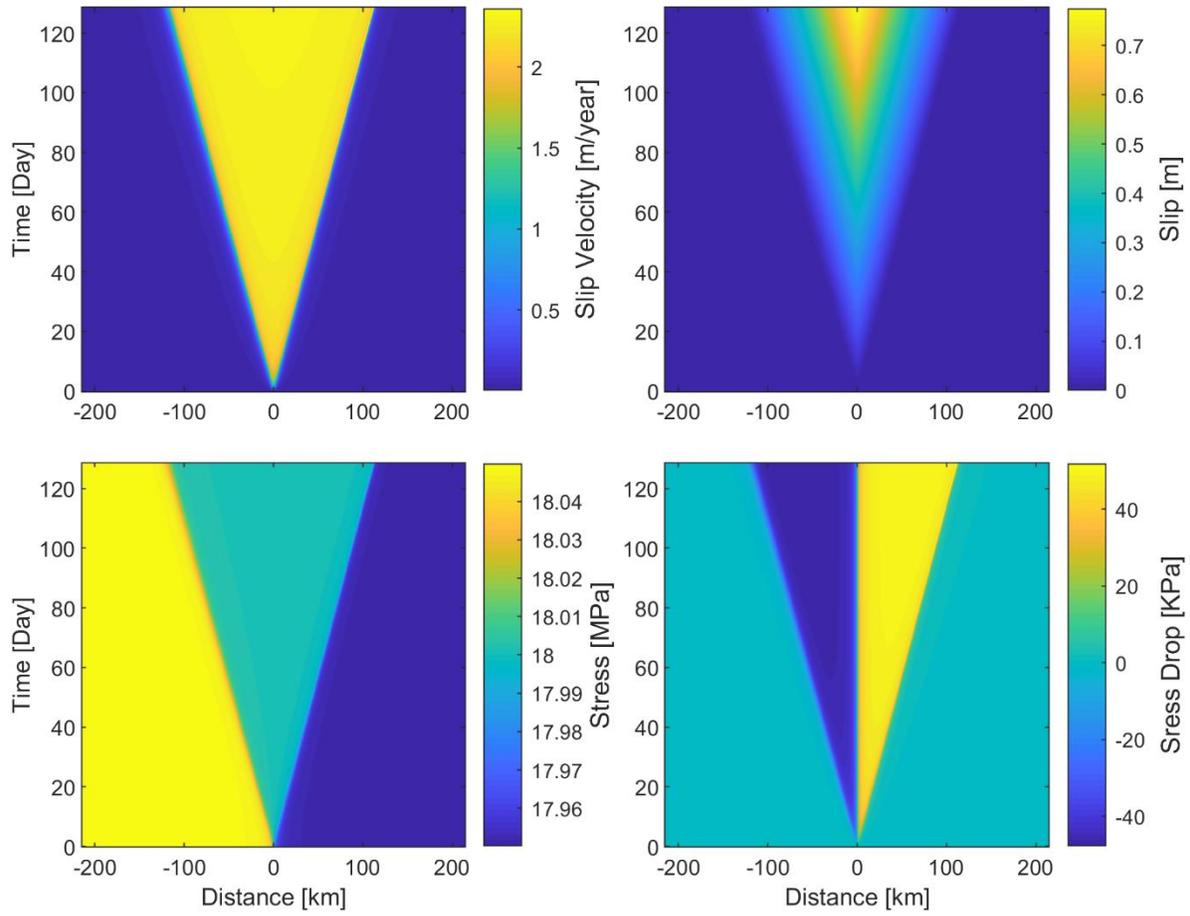

Figure 5. As in Fig. 4 (initial stress distribution of Eq. 12*a*), except for model conditions $\Sigma_{S0} = 0.18$ and $\delta\Sigma = 5 \cdot 10^{-4}$ that give a SSE. The rupture propagation velocities are $V^- \approx V^+ \approx 0.88$ km/day.

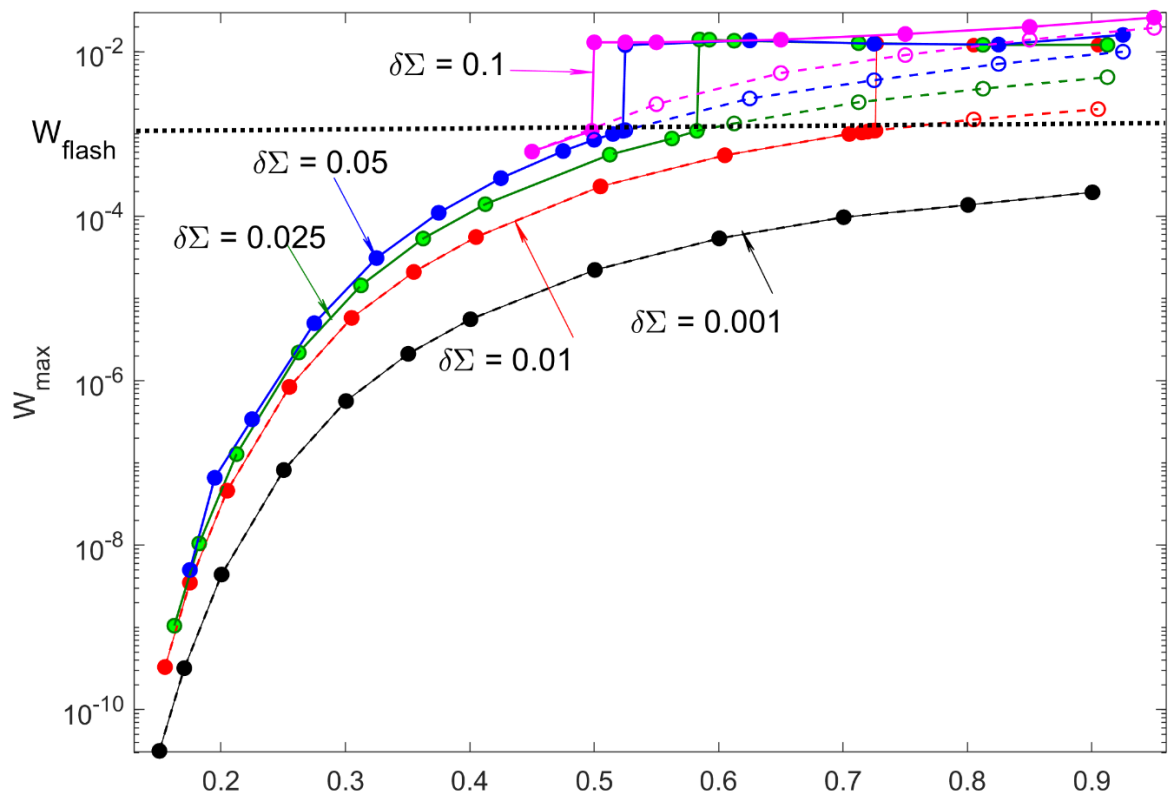
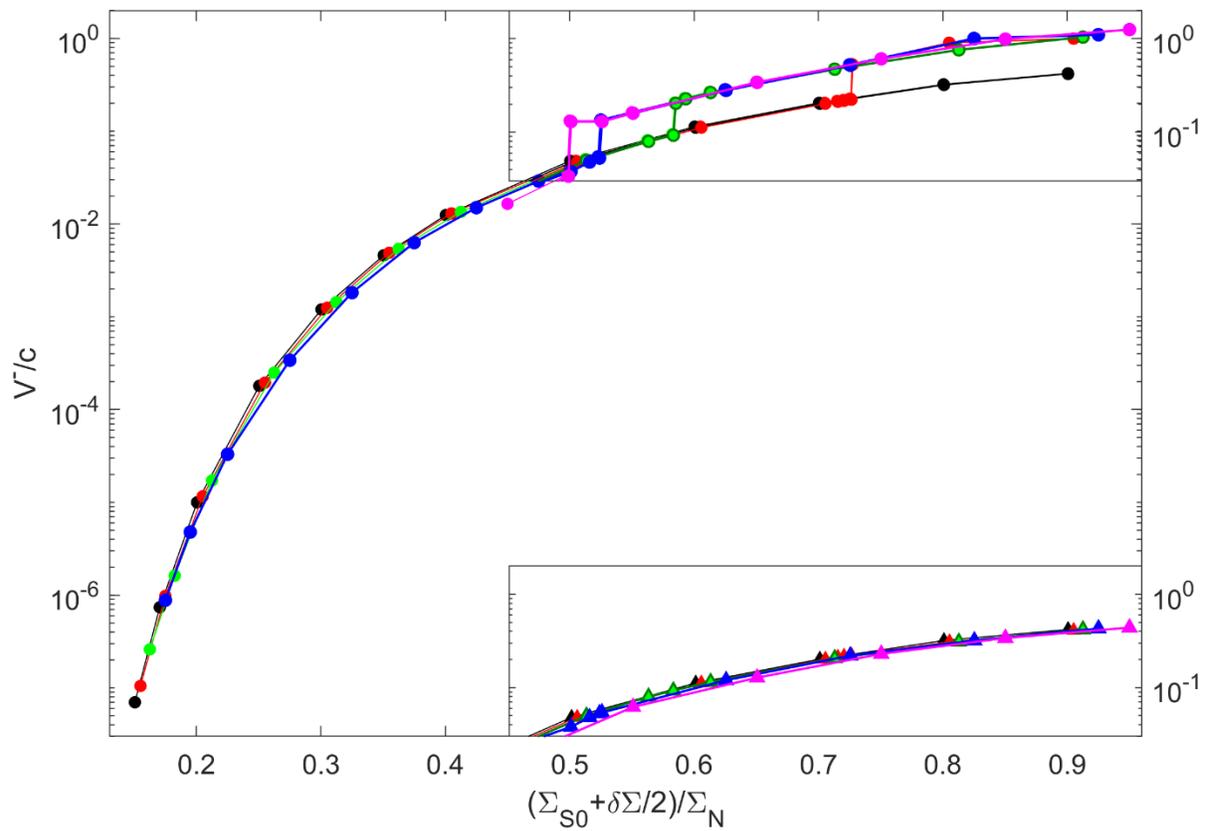

Figure 6. The dynamics of the variables that characterize a rupture depend critically on the initiating shear to normal stress ratio, as illustrated in Figs. 3–5. In (**a**), the maximum (dimensionless) slip velocity, $W_{max}$, occurring within a rupture is plotted as a function of the peak shear stress, $\Sigma_{max} = \Sigma_{S0} + \delta\Sigma/2$, relative to the normal stress $\Sigma_N$, for different values of $\delta\Sigma$. Solutions of the model Eqs. (5-12) were obtained using the initial stress distribution of Eq. 12*a* together with values listed in Table 1. Similar results are obtained for the distributions given in Eqs. 12 (b) and (c). The value for $W_{max}$ can be found from the value at the rupture initiation point $x = 0$ soon after the rupture occurs and at any subsequent times (see text). Abrupt jumps from the dashed to the solid curves result from the change in rate-state friction, Eq. (10), to flash heating, Eq. (11), when $W \geq W_{flash} = 1.1 \times 10^{-3}$ (0.1 m/s in physical units). Since regular EQs occur only if $W_{max}$ is larger than $1.1 \times 10^{-2}$ ($\geq 1$ m/s in physical units), we find that the reduction of friction due to flash heating is crucial for initiating EQ events. Otherwise, $W_{max}$ never attains the critical value, as shown in the dotted curves in (a), which were obtained using only the rate-state friction law. Rate-state friction using experimental values for *a* and *b* (see Table 1), can give $W_{max} \geq 10^{-2}$ only for unrealistically large values of the shear to normal stress (greater than ~ 0.9). The rupture propagation velocity $V^-$ is also plotted in (**b**), where the curves in the bottom right rectangle show the results of not including flash heating effects.

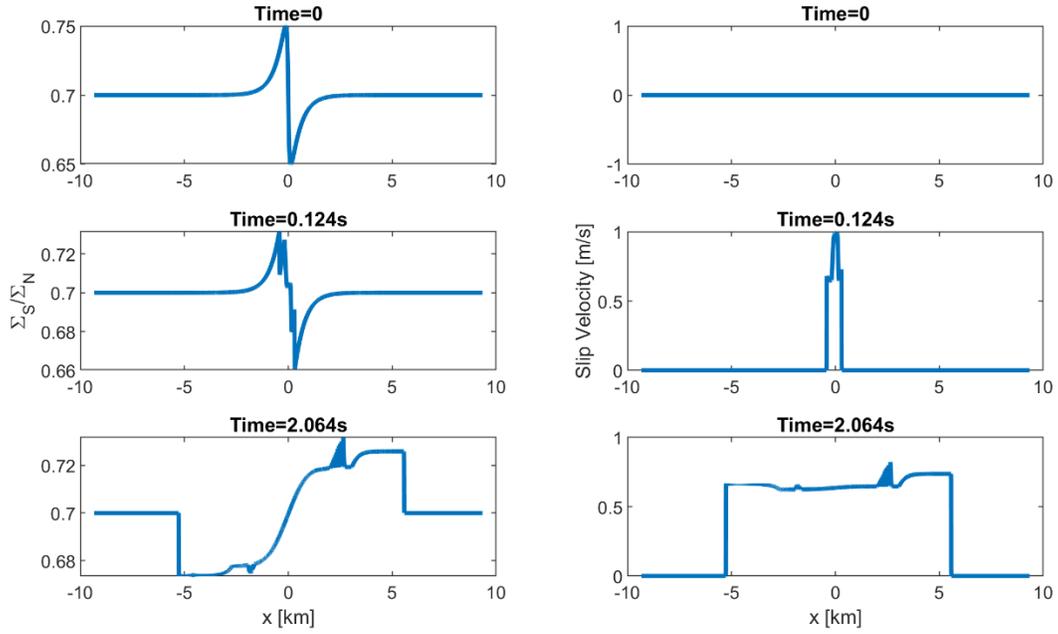

Figure 7. As in Fig. 3, except the initial stress distribution is given by Eq. (12b) (see also Fig. 2b). The example values $\Sigma_{S0} = 0.7$ and $\delta\Sigma = 0.17$ give results that are typical for regular EQs and produce a crack-like rupture as defined in section III B and seen more clearly in Fig. 8.

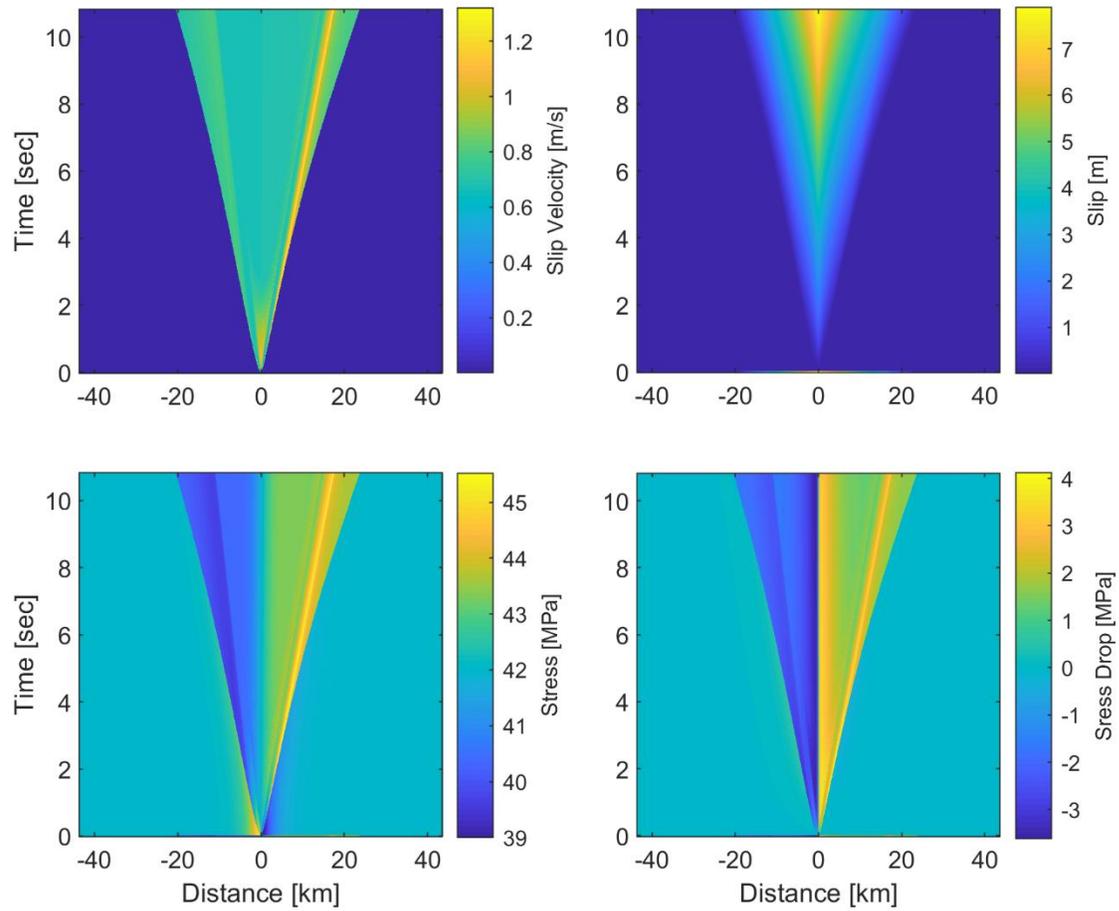

Figure 8. As in Fig. 7 (initial stress distribution of Eq. 12*b*), showing further detail for the spatial-temporal distribution of shear stress, $\Sigma_S$, and slip velocity, $W$, as well as stress drop, $\delta\Sigma$, and slip, $S$, for a regular, crack-like rupture.

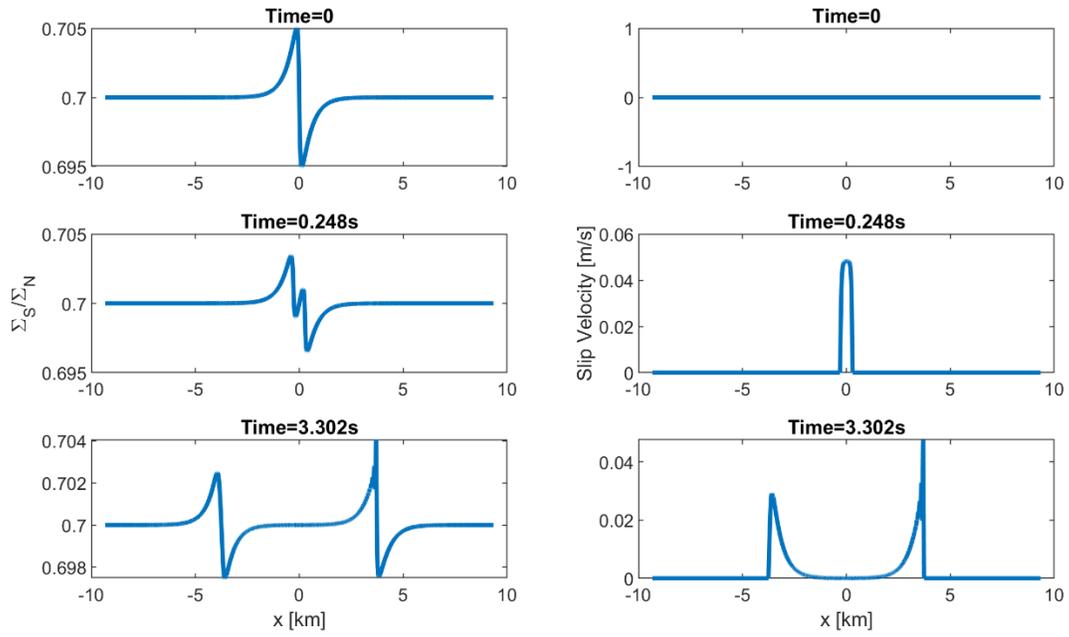

Figure 9. Similar to Fig. 7 (initial stress distribution of Eq. 12*b*), but using values $\Sigma_{s0} = 0.7$ and $\delta\Sigma = 0.017$ that result in a SSE. Note the reduction in $\delta\Sigma$ compared to the value used in Fig. 7. In this case, the rupture expands in both directions by self-healing (Heaton-like) pulses.

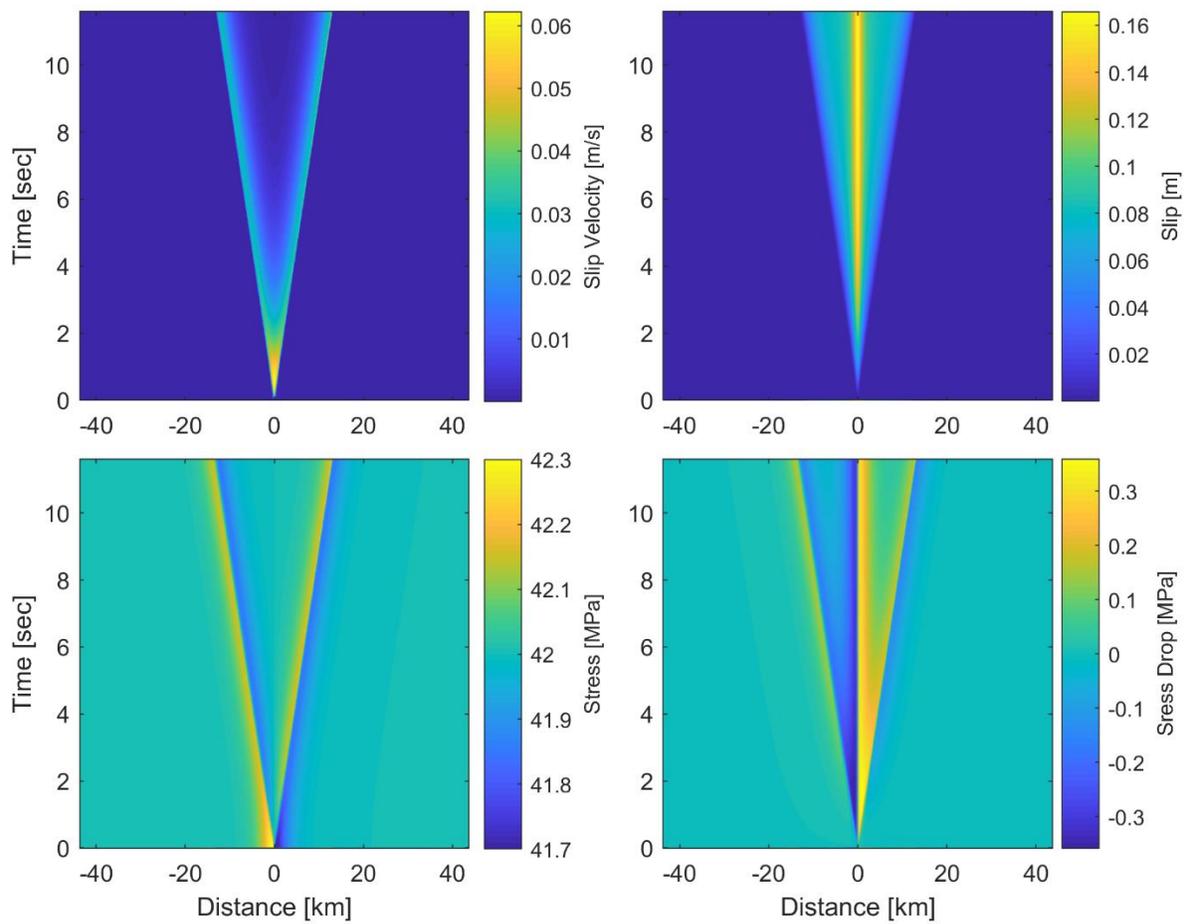

Figure 10. As in Fig. 9, showing further detail for the spatial-temporal distribution of shear stress, $\Sigma_S$, and slip velocity, $W$, as well as stress drop, $\delta\Sigma$, and slip, $S$. The slip velocity clearly shows the pulse-like nature of the rupture.

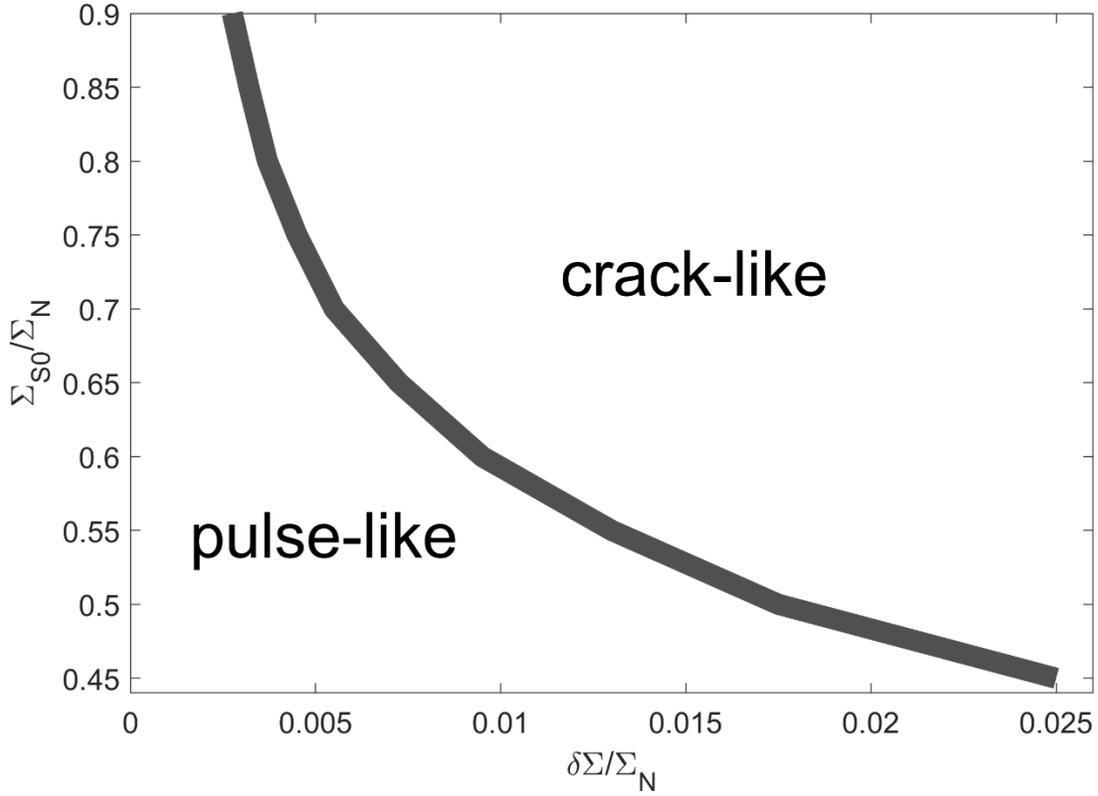

Figure 11. Values of the average shear stress, $\Sigma_{S0}$, and $\delta\Sigma$ (maximum minus minimum stress), normalized to the normal stress, $\Sigma_N$, that determine whether a rupture event is crack-like or pulse-like. Simulations were performed over wide ranges of $\delta\Sigma_{S0}$ and $\delta\Sigma$ for the initial stress distributions given by Eqs. 12. For the localized initial stress distributions considered here, if $W_{max} \geq W_{flash}$, the rupture is crack-like, while the rupture is pulse-like for $W_{max} < W_{flash}$. For a non-localized step-like stress distribution, ruptures are crack-like for all values of $\Sigma_{S0}$ and $\delta\Sigma$. See Eq. 12 and Fig. 2 for the stress profiles used in the simulations.

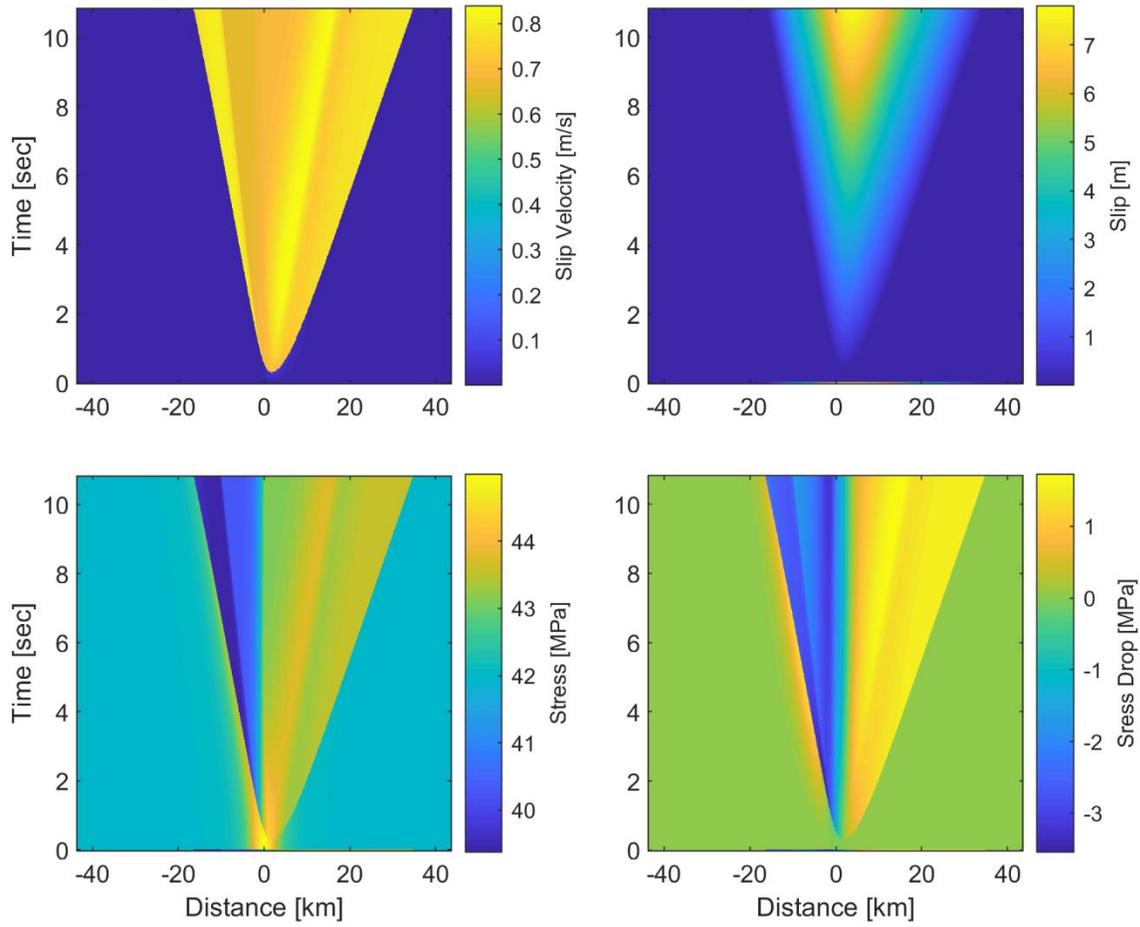

Figure 12. Spatial-temporal distribution of shear stress, $\Sigma_S$, slip velocity, $W$, stress drop, $\delta\Sigma$, and slip, $S$, for the initial stress distribution of Eq. 12*c* (see also Fig. 2*c*). The example values $\Sigma_{S0} = 0.7$ and $\delta\Sigma = 0.05$ give results that are typical of regular EQs, which are characterized by crack-like ruptures. However, the position of the maximum value for *S* moves to the right in time, in contrast to results for the initial distributions of Eqs. 12*a* and 12*b* in which the maximum remains centered at the initiation point $x = 0$.

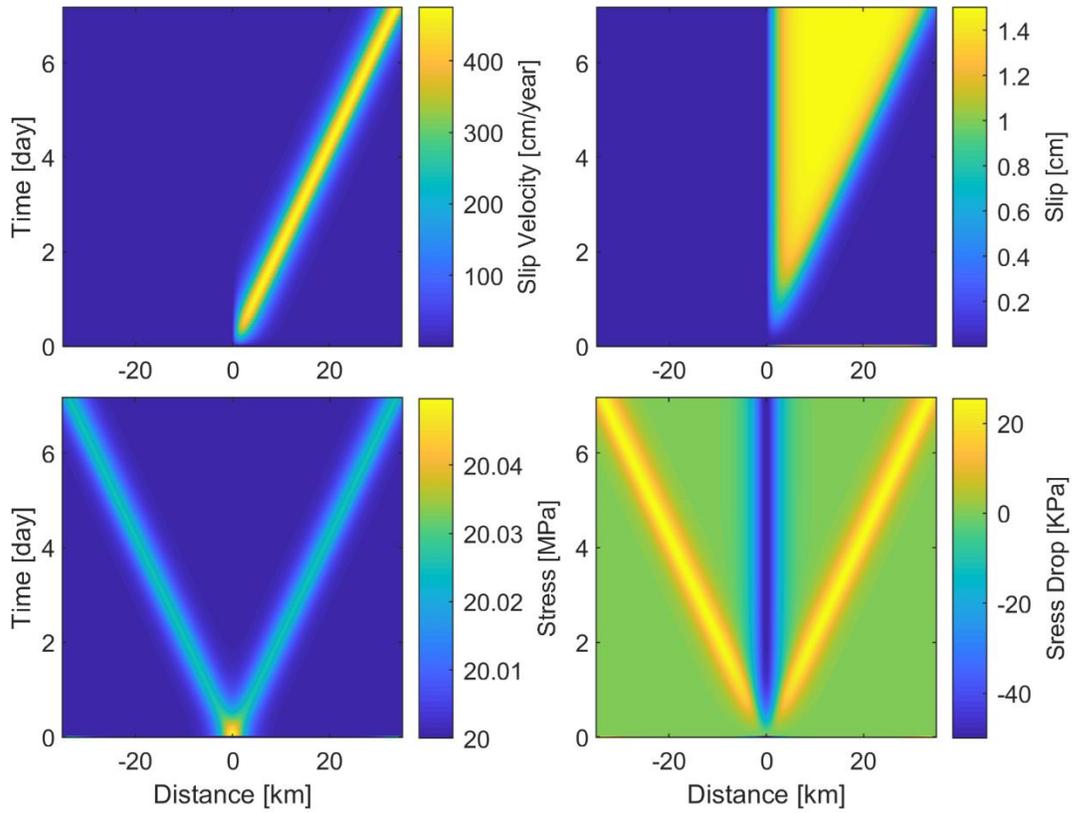

Figure 13. As in Fig. 12, but for conditions $\Sigma_{s0} = 0.2$ and $\delta\Sigma = 5 \cdot 10^{-4}$ that give a SSE and a pulse-like rupture. For comparison, the values of initial stress, Eq. 12b, used in Fig. 10, also produce a SSE and generate a pulse-like rupture. However, in the present case, the rupture now expands in only one direction, as shown by the spatial-temporal behavior of the slip and slip velocity. There are still two stress pulses, as for the initial condition of Eq. 12b, but the stress pulse at $x < 0$ is *not* accompanied by a slip pulse.